\documentclass[twocolumn, resetfootnote]{aastex702}

\usepackage{amsmath}
\usepackage{microtype}

\newcommand{\comment}[1]{}
\newcommand{\lcdm}{$\Lambda$CDM}
\newcommand{\tabnote}[1]{\multicolumn{8}{@{}p{\dimexpr\textwidth-20pt\relax}@{}}{#1}}

\begin{document}

\title{
    A First Measurement of Circumgalactic Dust Reddening\\
    from Only 4.6\,deg$^2$ of the Rubin Observatory's Data~Preview~1
}

\author[0000-0002-2495-3514]{John Franklin Crenshaw}
\affiliation{Kavli Institute of Particle Astrophysics and Cosmology, USA}
\affiliation{Department of Physics, Stanford University, Stanford, CA 94305-4085, USA}
\affiliation{SLAC National Accelerator Laboratory, 2575 Sand Hill Road, Menlo Park, CA 94025, USA}
\email[show]{jfcren@stanford.edu}

\author[0000-0001-7961-9735]{Matthew McQuinn}
\affiliation{Department of Astronomy, University of Washington, Seattle, WA 98195, USA}
\email{mcquinn@uw.edu}

\author[0000-0002-0355-0134]{Jessica K. Werk}
\affiliation{Department of Astronomy, University of Washington, Seattle, WA 98195, USA}
\email{jwerk@uw.edu}

\begin{abstract}

We present the first measurement of circumgalactic dust reddening from the Vera C. Rubin Observatory, using only 4.6\,deg$^2$ of ComCam Data Preview 1 --- roughly $0.03\%$ of the final LSST footprint.
Using photometric redshifts, we stack background-galaxy colors around foreground-galaxy positions and detect a chromatic reddening profile from $r_\perp \simeq 10$\,kpc to $1$\,Mpc.
Interpreting average $E(g-z)$ with a Milky Way extinction curve, we find
$A_V = (1.3 \pm 0.4) \times 10^{-1} (r_\perp / 20\,\mathrm{kpc})^{-1.8 \pm 0.4}$ within $120$\,kpc.
The amplitude and radial dependence agree with earlier Sloan Digital Sky Survey (SDSS), KiDS, and Dark Energy Survey (DES) results despite the $\sim1000\times$ smaller survey area and a foreground sample extending 3--6\,mag fainter and 1--2\,dex lower in stellar mass.
The innermost 10--15\,kpc bin reaches $A_V \simeq 0.3$\,mag, comparable to high-latitude extinction through the Milky Way disk near the solar circle; the steep power-law slope implies a dust distribution that does not simply trace the halo-gas profile.
Splitting by rest-frame $g-r$ shows stronger extinction around red foreground galaxies (rest-frame $g-r > 0.5$), although the blue subsample is too noisy to establish a significant color dependence.
This red sample, with median halo mass $5 \times 10^{11}\,M_\odot$, shows substantially more reddening within 50\,kpc than previously measured around more massive LRGs and implies a dust-to-stellar-mass ratio of $\sim 2\%$, nearly saturating the dust budget allowed by stellar metal yields.
These pathfinder data demonstrate LSST's promise for high-precision galaxy--dust measurements across galaxy mass, environment, and redshift.

\end{abstract}

\section{Introduction}
\label{sec:intro}

The circumgalactic medium (CGM) encompasses the diffuse gas, metals, and dust surrounding galaxies, extending through their dark matter halos and connecting the interstellar medium (ISM) to the intergalactic medium (IGM).
As the interface between galaxies and the cosmic web, the CGM regulates galaxy evolution by mediating gas accretion, recycling stellar material, hosting metal-enriched outflows driven by stellar and AGN feedback, and providing a reservoir for future star formation \citep[e.g.,][]{tumlinson2017,giguere2023}.
Absorption-line studies with background quasars and galaxies show that the CGM is multiphase, with cool photoionized gas, warm metal-bearing gas, and hot virialized plasma all contributing to the halo baryon budget \citep{maller2004,tumlinson2011,werk2014,faerman2017}.

Dust is another tracer of this baryon cycle.
Interstellar grains can be grown in the dense material around evolved stars and supernovae, or even in more tenuous interstellar gas.
They can be ejected beyond the galaxy into its dark matter halo by radiation pressure and feedback, where they may be destroyed by shocks and thermal sputtering in the virial plasma \citep{chiao1972,chiao1973,draine1979,draine1979destruction,dwek1992,dwek1998,draine2003,galliano2018}.
Dust abundance therefore probes enrichment, outflows, and CGM conditions, complementing absorption-line measurements of the gas-phase metals \citep{otsuki2024}.

Multiple lines of evidence indicate the CGM contains appreciable dust.
Quasar absorption systems, especially Mg~{\sc ii} absorbers, redden background quasars and exhibit dust-to-gas ratios comparable to those of galaxies \citep{york2006,menard2008,menard2012}.
Nearby edge-on galaxies reveal extraplanar dust in optical absorption \citep{howk1999} and far-infrared halo emission \citep{yoon2021}, while SOFIA/HAWC+ polarimetry detects emission from magnetically aligned grains in starburst outflows several kpc above the disk \citep{jones2019sofia,lopezrodriguez2023}.
Together, these observations demonstrate that feedback, winds, and tidal interactions can transport dust out of galaxies.

Statistical reddening measurements extend this picture to larger scales by averaging over many foreground--background pairs to probe the projected galaxy--dust correlation.
An early implementation of this approach came from \citet{zaritsky1994}, who used the colors of background galaxies seen through the halos of NGC~2835 and NGC~3521 to find preliminary evidence for enhanced reddening at projected radii of $\sim60$\,kpc relative to control fields at $\sim220$\,kpc.
Wide-field surveys later made this approach far more powerful.
\citet{menard2010} measured the reddening of $\simeq 85,000$ background quasars around foreground galaxies in $\simeq 3800\,\text{deg}^2$ of the Sloan Digital Sky Survey (SDSS), detecting a color excess from $\simeq 20$\,kpc to several Mpc, and \citet{peek2015} performed a similar analysis using SDSS ``standard crayons,'' a galaxy sample selected for uniform colors.
These SDSS measurements found profiles consistent with $\langle E(B-V) \rangle \propto r^{-0.8}$, with \citet{peek2015} finding evidence for enhanced reddening on inner CGM scales.
The inferred dust masses of $\sim 10^7$--$10^8 \, M_\odot$ around low-redshift $L_*$ galaxies suggest that roughly half of their dust lies in the CGM.
The amplitude and wavelength dependence further imply that dust, including small grains vulnerable to sputtering, survives transport through winds to the CGM.

The large-scale distribution of dust also connects galaxy-formation physics to precision cosmology, as stellar and AGN feedback redistribute baryons on nonlinear scales, possibly out to several virial radii \citep{chisari2019}.
A range of observables now constrains this redistribution, with each probe emphasizing different components of the baryon cycle and carrying different modeling sensitivities.
Weak lensing surveys, which trace total mass, constrain the baryonic suppression of small-scale clustering, although the inferred feedback strength depends on modeling choices \citep{arico2023,grandon2024}.
Thermal and kinetic Sunyaev-Zel'dovich (tSZ/kSZ) measurements trace hot and ionized gas through pressure and optical depth \citep{amodeo2021,pandey2022,riedguachalla2025}, while fast radio burst (FRB) dispersion measures probe the free-electron column density \citep{mcquinn2014, macquart2020, sharma2026}.
Because dust traces enriched material in a cooler, potentially denser phase, galaxy--dust reddening adds a complementary view of feedback-driven baryon redistribution.

Intergalactic dust also dims and reddens background sources, coupling to magnification measurements, photometric-redshift inference, galaxy clustering, and Type~Ia supernova distances \citep{aguirre1999,mortsell2003,menard2010,menard2010sn,goobar2018}.
Indeed, uncertainties in extragalactic dust extinction are quoted as the second largest source of systematic error in recent Type~Ia supernova analyses \citep{rubin2025}.

Simulations broadly reproduce the amplitude and radial scaling of these dust profiles, but remain discrepant in their evolution with redshift and stellar mass \citep{mckinnon2017,aoyama2018,qiao2024,parente2026}, motivating empirical constraints across a wide range of galaxy properties and cosmic epochs.
These uncertainties extend to the transport physics itself, with recent simulations suggesting grain growth in entrained cool gas may be needed to maintain high dust-to-gas ratios \citep{hu2026}.
Galaxy-formation models also remain divided on how the kinematics and thermodynamics of the CGM are shaped by cosmic rays (CRs), which can drive comparatively cool, smooth, and mass-loaded outflows \citep{Girichidis2018,Hopkins2021,RodriguezMontero2024}.
Such environments would be capable of transporting and maintaining larger dust masses in the CGM \citep{Salem2016,Ji2020,Butsky2020,Butsky2022,martin-alvarez2026}.
Empirical galaxy--dust correlations therefore constrain feedback-driven enrichment, CR-sensitive CGM thermodynamics, and dust-related systematics in imaging surveys.

The SDSS results have since been extended to more recent galaxy surveys.
\citet{genc2025} used the bright ($r<20$) sample of the 1000\,deg$^2$ Kilo-Degree Survey (KiDS) as foreground lenses, jointly fitting magnitude shifts and galaxy--galaxy lensing to infer a dust mass of $\sim 7 \times 10^7 \, M_\odot$ for halos of $\sim 10^{12.3} \, M_\odot$.
\citet{mccleary2026} used Luminous Red Galaxies (LRGs) from the 5000\,deg$^2$ Dark Energy Survey (DES) as background tracers, finding $A_V \propto r^{-0.8}$ beyond $50\,h^{-1}$\,kpc around both their WISExSuperCOSMOS and LRG foreground samples.
Together with SDSS, these analyses span a range of foreground selections, background samples, and estimators, making them useful points of comparison.
Notably, they are all in qualitative agreement, except for the LRG measurements of \citet{mccleary2026}, which suggest these galaxies harbor very little dust in their inner CGM.
These studies have required large survey areas because the dust signal is subtle: the typical color excess is only $\sim 10^{-3}$\,mag at Mpc scales, far below the intrinsic color scatter of galaxies.

\begin{table*}[t]
    \caption{
        Summary of Rubin DP1 fields used in this analysis.
        Statistics are computed across pixels that pass our quality cuts.
        $r_5$ is the 5$\sigma$ depth in the $r$ band.
        \label{tab:dp1}
    }
    \centering
    \begin{tabular}{lccccc}
        \hline\hline
                                & ECDFS    & EDFS     & SV~95~-25 & SV~38~7 & Total \\
        \hline
        Area (deg$^2$)          & 1.1      & 0.90     & 0.85      & 1.7     & 4.6   \\
        Median $r_5$            & 26.1     & 25.7     & 25.8      & 24.5    & 25.2  \\
        Foreground galaxies     & 5859     & 5043     & 5659      & 2012    & 18573 \\
        Background galaxies     & 7536     & 5042     & 5431      & 1701    & 19710 \\
        Fraction of fg--bg pairs & 38\%     & 26\%     & 34\%      & 2\%     & \\
        Photo-$z$ bands         & $ugrizy$ & $ugrizy$ & $ugrizy$  & $griz$  & \\
        \hline
    \end{tabular}
\end{table*}

The Vera C. Rubin Observatory's Legacy Survey of Space and Time \citep[LSST;][]{ivezic2019} promises to transform this landscape.
LSST will image $\sim18,000$ deg$^2$ in 6 bands ($ugrizy$) to unprecedented depth over 10 years, detecting $\sim20$ billion galaxies with high-quality photometry.
This combination of area, depth, and multiband coverage will greatly expand the statistical power and scope of circumgalactic dust studies, enabling measurements across redshift and as a function of foreground-galaxy properties such as color, stellar mass, star-formation rate, and orientation relative to the disk.

Here we present the first constraints on circumgalactic dust from the Rubin Observatory.
\textit{Using only 4.6\,deg$^2$} from Data Preview 1 (DP1; \citealt{dp1}), a commissioning pathfinder obtained with Rubin's Commissioning Camera (ComCam), we show Rubin-quality photometry already yields a signal consistent with previous wide-area surveys, while probing fainter foreground galaxies and lower stellar masses.
This result previews LSST's ability to characterize circumgalactic dust across cosmic time. 

Section~\ref{sec:data} describes the data and our sample selection, which combines two independent photo-$z$ estimates, the template-based \texttt{LePhare} and the machine-learning \texttt{FlexZBoost}, to define cleanly separated foreground ($0.2 < z_\text{phot} < 0.5$) and background ($0.7 < z_\text{phot} < 1.4$) samples.
Section~\ref{sec:methods} presents the stacking method, in which we measure the mean color excess of background galaxies in bins of projected separation from the foreground galaxies, using ``flipped'' stacks and random catalogs to remove systematics.
Section~\ref{sec:results} presents our results, Section~\ref{sec:discussion} interprets them and compares to previous work, and Section~\ref{sec:conclusion} concludes.
We assume \citet{planck2018} \lcdm{} cosmology throughout.

\section{Data}
\label{sec:data}

\begin{figure*}[t]
    \centering
    \includegraphics{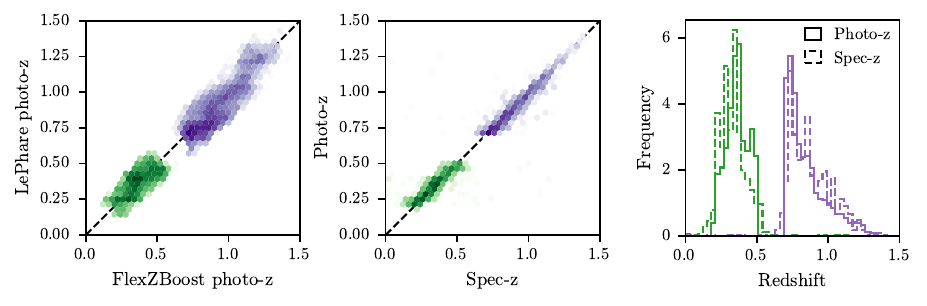}
    \caption{
        Photometric and spectroscopic redshift distributions for the foreground (green) and background (purple) galaxy samples.
        Left: \texttt{LePhare} vs \texttt{FlexZBoost} photo-$z$ estimates.
        We remove galaxies for which the two estimates differ by more than $0.1 (1 + z_\text{phot})$.
        For the remainder, we assign the inverse-variance-weighted mean of the two methods for our photo-$z$ estimate.
        Middle: photo-$z$ vs spec-$z$ for galaxies with spectroscopy.
        Right: photo-$z$ and spec-$z$ histograms.
        These distributions demonstrate the robustness of our photo-$z$ selection and estimates, as well as the low cross contamination between the two samples.
    }
    \label{fig:photoz}
\end{figure*}

\subsection{Rubin DP1}
\label{sec:DP1}

Rubin's ComCam \citep{comcam2024,2020SPIE11447E..0LS,2022SPIE12184E..0JS} is a 144 megapixel pathfinder instrument covering roughly 5\% of the LSSTCam focal plane, providing an $\approx0.5$\,deg$^2$ field of view.
Its focal plane comprises nine high-performing spare LSSTCam CCDs, while the telescope, readout systems, and data pipelines are the same as those used for LSSTCam, enabling end-to-end testing of the observatory.

ComCam observed on sky from October 24 to December 11, 2024, producing the DP1 dataset described by \citet{dp1}.
We use RA, Dec, and $ugrizy$ AB magnitudes from the coadded object catalog\footnote{Schema: \url{https://sdm-schemas.lsst.io/dp1.html}}.
For fluxes in individual bands, we use \texttt{cModel} fluxes, which come from a two-component bulge+disk model fit; for colors, we use \texttt{gaap1p0} fluxes, which use a 1\,arcsec Gaussian aperture after homogenizing the point-spread function (PSF) across bands, providing more consistent colors.

To assemble an approximately magnitude-limited galaxy sample, we select extended objects (\texttt{refExtendedness == 1}) with $r < 24$.
We pixelize the footprint with healpix \texttt{NSIDE=1024}, estimate each pixel's 5$\sigma$ depth in each band from the median \texttt{cModel} magnitude error, remove pixels with $r$-band 5$\sigma$ depth shallower than 24, and then remove the shallowest remaining 5\% of pixels in $giz$.
The final area is $\approx 4.6$\,deg$^2$ across four DP1 fields (ECDFS, EDFS, SV~38~7, SV~95~-25; Table~\ref{tab:dp1}).

\subsection{Known issues with DP1}
\label{sec:dp1-issues}

While the DP1 dataset provides a uniquely faithful preview of LSST, several commissioning-era issues degrade its quality relative to the data expected from LSST \citep{dp1}.
These include high ComCam read noise, elevated charge transfer inefficiency, elevated operating temperature and associated $g$-band red leak, evolving active optics and thermal control, and uncontrolled dome turbulence.
DP1 is therefore a conservative preview of LSST performance.

The photometric errors in the DP1 $u$ and $y$ bands are also significantly underestimated, and their depths are overestimated \citep{sitcomtn-154}.
To avoid these calibration complications, we remove sources with $z_\text{phot}<0.2$ (Section~\ref{sec:photoz}) and compute galaxy colors only from $griz$.
Note that the photo-$z$ estimates for three of our four fields (ECDFS, EDFS, SV~95~-25) \emph{were} computed using $u$ and $y$; however, our fourth field (SV~38~7) lacks any exposures in these bands.

\begin{figure*}[t]
    \centering
    \includegraphics{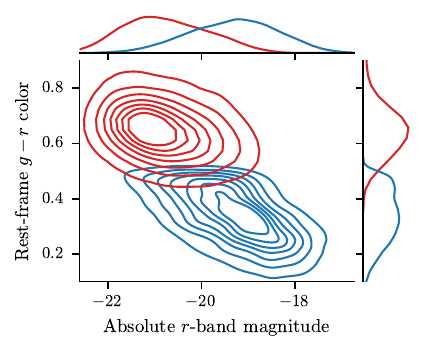}
    \includegraphics{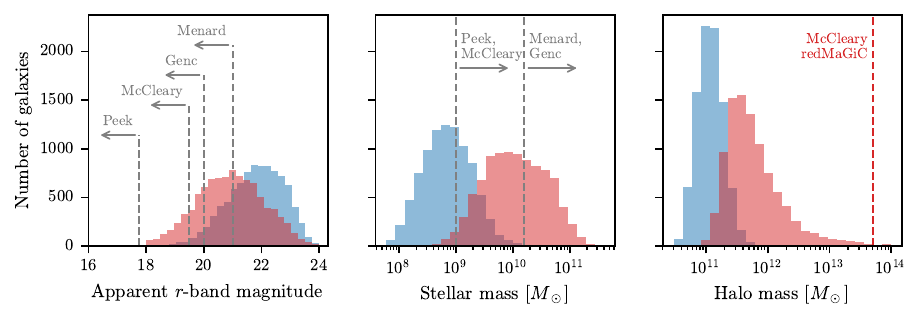}
    \caption{
        Properties of the foreground sample.
        Upper: rest-frame color-magnitude diagram.
        We use a cut of $g - r = 0.5$ to split the foreground sample into red and blue subsamples.
        Left: observed-frame apparent $r$-band magnitudes.
        Dashed lines and arrows mark magnitude limits from previous studies.
        Middle: estimated stellar masses.
        The red sample is more massive than the blue sample.
        Dashed lines and arrows mark approximate lower mass limits from previous studies.
        Right: estimated halo masses.
        The limits labeled McCleary are for the WISExSuperCOSMOS sample of \citet{mccleary2026} in the left and middle panels, and for their redMaGiC LRG sample in the right panel \citep{zacharegkas2022}.
    }
    \label{fig:foreground}
\end{figure*}

\subsection{Photo-$z$ Selection}
\label{sec:photoz}

We select foreground and background galaxies using DP1 photo-$z$ estimates from the RAIL team \citep{zhang2025b,rail}.
For both the \texttt{LePhare} \citep{lephare1,lephare2} template photo-$z$ and the \texttt{FlexZBoost} \citep{flexzboost} machine-learning photo-$z$, we define $\sigma_\text{pz} = 0.5 \,(z_{84} - z_{16})$, where $z_{84}$ and $z_{16}$ are the 84$^{\rm th}$ and 16$^{\rm th}$ percentiles of the photo-$z$ posterior.
In order for the galaxy to be included in our sample, we require\footnote{We anticipate this requirement also reduces contamination from blending; see Appendix~\ref{app:inner-bin}.} $\sigma_\text{pz} < 0.1$ for both estimates and $|z_\texttt{LePhare} - z_\texttt{FlexZBoost}| < 0.1 (1 + z_\text{phot})$.
For each galaxy's redshift, the estimated photometric redshift, $z_\text{phot}$, is the inverse-variance-weighted mean of the two photo-$z$ estimates.

We then select foreground galaxies with $0.2 < z_\text{phot} < 0.5$ and background galaxies with $0.7 < z_\text{phot} < 1.4$.
The intervening $\Delta z=0.2$ gap limits overlap and contamination, while the $z_\text{phot}>0.2$ and $z_\text{phot}<1.4$ cuts avoid regimes most affected by the DP1 $u$- and $y$-band issues by removing redshift ranges where these bands are most essential for localizing the Balmer break.
The low-redshift cut also limits the angular size subtended by fixed projected separations.
The resulting sample contains 18,573 foreground galaxies and 19,710 background galaxies.

We evaluate the resulting photo-$z$ quality in Fig.~\ref{fig:photoz} by comparing the estimates to the spectroscopic redshifts in the DP1 redshift training catalog.
This comparison demonstrates that our foreground and background samples have relatively accurate photo-$z$ estimates, with minimal overlap and contamination between the two samples.
Indeed, the contamination rate for our spectroscopically crossmatched galaxies is only 0.01\%, using $z=0.6$ as the classification boundary.
We also perform analysis variants changing the redshift gap between the foreground and background selection cuts by $\pm 0.1$ and confirm that the results are consistent with our fiducial analysis within the uncertainties (see Appendix~\ref{app:sensitivity}).

\subsection{Characterizing the foreground sample}

Because the stack measures dust around the selected foreground population, the stacked sample's luminosities, colors, and masses are central to the interpretation.
We estimate rest-frame magnitudes, colors, and stellar masses with \texttt{kcorrect} \citep{blanton2007}, using the expanded template set of \citet{pai2024}.
Model residuals are $\lesssim 1\%$ in $griz$.
We estimate halo masses by inverting the redshift-dependent stellar-to-halo mass relation of \citet{moster2013}, which was determined by abundance matching.
These masses are intended as approximate population descriptors rather than precise masses for individual galaxies.

While the statistical precision of the DP1 dataset is not sufficient to split across a wide range of properties, we do consider a split by galaxy color.
Figure~\ref{fig:foreground} shows a bimodal rest-frame $g - r$ distribution, motivating a split into red and blue subsamples at $g - r = 0.5$.
The blue subsample reaches fainter apparent magnitudes, as well as lower stellar and halo masses.

Relative to previous galaxy--dust measurements, the DP1 foreground selection is substantially deeper: it reaches $r<24$ and extends the stellar-mass range down to $\sim10^8 \, M_\odot$.
Because color also correlates with stellar mass, luminosity, dust attenuation, inclination, and environment, the split is not a pure star-forming/quiescent division, but this split provides a glimpse into how galaxy--dust correlations change across foreground populations.

\section{Methods}
\label{sec:methods}

We measure the average color excess from circumgalactic dust by stacking flux ratios of background galaxies around the positions of foreground galaxies, and expressing the stacked ratios as colors in magnitude units.

For each foreground--background pair $(f, b)$, we compute the physical transverse separation of $r_\perp = D_A(z_f^\text{phot})\,\theta_{fb}$, where $D_A$ is the angular diameter distance to the foreground photo-$z$ and $\theta_{fb}$ is the pair angular separation on the sky.
We bin pairs into four bins from $0.01$ to $1$\,Mpc.
For a band pair $x-y$, we compute the background-galaxy flux ratio
\begin{align}
    \rho_b^{xy} = f_{x,b} / f_{y,b},
\end{align}
using the fluxes corresponding to the \texttt{gaap1p0} photometry.
In each radial bin $i$, we form a weighted mean of these ratios,
\begin{align}
    \bar{\rho}^{\,xy}_i &=
        \frac{\sum_{(f,b) \in i}\, w^{xy}_{b,i}\,\rho_b^{xy}}{\sum_{(f,b) \in i}\, w^{xy}_{b,i}},
\end{align}
where the inverse-variance weights include a floor set by the intrinsic variance within the radial bin:
\begin{align}
    w^{xy}_{b,i} &=
        \frac{1}{\sigma^2_{\rho_b} + (s^{xy}_{\rho,i})^2}.
\end{align}
$\sigma_{\rho_b}$ is the propagated photometric uncertainty on the flux ratio and $(s^{xy}_{\rho,i})^2$ is the empirical variance of $\rho^{xy}_b$ in the radial bin.
This floor captures intrinsic background-galaxy color scatter, which dominates the propagated photometric uncertainty for most pairs.
We then convert the stacked ratio to a magnitude color,
\begin{align}
    \bar{c}^{\,xy}_i = -2.5 \log_{10} \bar{\rho}^{\,xy}_i.
    \label{eq:raw-estimator}
\end{align}
We use $g-z$ as our primary dust tracer, but also compute stacked ratios for $g-r,\, r-i,\, i-z$ to probe the chromaticity of the inferred reddening.

The estimator in Eq.~\ref{eq:raw-estimator} may be contaminated by large-scale variations in photometric calibration, image depth and quality, and Galactic extinction, as well as photometric errors like background oversubtraction.
To correct for these residual systematics, we compute a ``flipped'' stack $\bar{c}^\text{\,flip}_i$ by applying the same estimator to \emph{foreground} galaxy colors around the positions of background galaxies, while still using the foreground photo-$z$ to compute $r_\perp$.
The flipped stack has no physical circumgalactic-reddening signal, but inherits the same selection and systematic structure, so we adopt
\begin{align}
    \Delta \bar c_i \equiv \bar{c}_i - \bar{c}^\text{\,flip}_i
\end{align}
as an initial systematics-corrected estimator \citep[cf.][]{peek2015}.

We further correct for the finite and nonuniform survey footprint using random catalogs drawn from the same selected HEALPix footprint and jackknife regions as the galaxy sample.
These randoms encode the survey geometry and are matched to the depth and foreground redshift distribution of the corresponding real samples.
We measure the same forward-minus-flipped estimator around random positions and subtract it from the data,
\begin{align}
    \Delta \bar c^\mathrm{corr}_i
    \equiv
    \Delta \bar c_i - \Delta \bar c^\mathrm{rand}_i.
\end{align}
The detailed random-catalog construction, including the depth-matching weights, is described in Appendix~\ref{app:randoms}.

We obtain the final estimator by subtracting the average value of this same random-corrected quantity in a reference annulus at $2 < r_\perp\,/\,\text{Mpc} < 4$:
\begin{align}
    E_i(c) =
    \Delta \bar c^\text{\,corr}_i
    - \Delta \bar c^\text{\,corr}_\text{ref}.
\end{align}
This removes the intrinsic mean galaxy color and any constant offset between the regular and flipped stacks.
This latter effect is especially important for removing edge effects where the depth rolls off at the edge of the fields.

We estimate the covariance of $E_i(c)$ with a jackknife (using the ``match'' weighting of \citealt{mohammad2022}) over 12 spatial regions, splitting each of the four DP1 fields into three contiguous, roughly equal-area subregions.
This small number is forced by the DP1 footprint relative to the largest scale in our estimator.
The measurement noise floor is therefore set by jackknife covariance precision rather than per-pair photometric scatter, which is orders of magnitude smaller.
Future LSST releases should improve the signal-to-noise by almost 2 orders of magnitude by providing $4000 \times$ more area.

The dominant uncertainty in $r_\perp$ is the foreground photo-$z$.
For our foreground sample (Section~\ref{sec:photoz}), the typical scatter $\sigma_z \lesssim 0.1$ corresponds to $\lesssim 15\%$ uncertainty in $r_\perp$, which is small compared to our wide logarithmic bins.

Appendix~\ref{app:sensitivity} demonstrates the stability of our estimator under a variety of analysis choices, indicating that our analysis does not rely on the precise random-catalog weighting, the exact placement of the reference annulus, the adopted width of the photo-$z$ gap between the foreground and background samples, the choice between flux ratios and magnitude differences, or the inclusion of the DP1 SV~38~7 field without $u$- and $y$-band data.

\section{Results}
\label{sec:results}

\begin{figure}
    \centering
    \includegraphics[width=\linewidth]{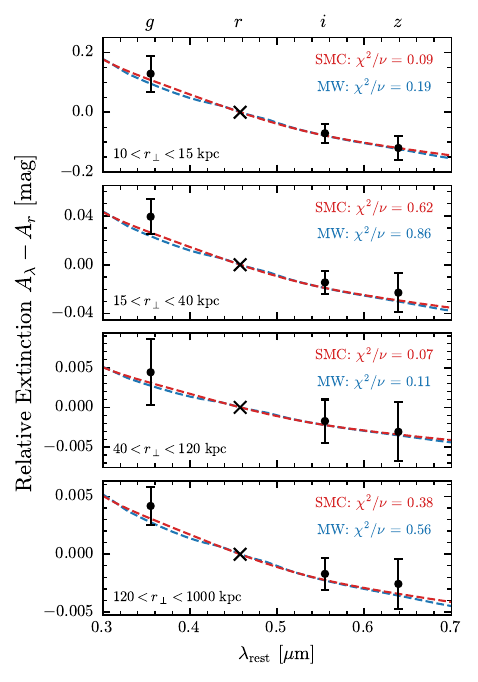}
    \caption{
        Observed extinction relative to the $r$ band, compared with extinction curves for the Milky~Way (blue; \citealt{gordon2023} with $R_V=3.1$) and the SMC bar (red; \citealt{gordon2003}).
        Each panel shows one radial bin.
        The MW and SMC normalizations are fit independently in each bin; the corresponding $\chi^2$ per degree of freedom is shown in the upper right.
    }
    \label{fig:chromaticity}
\end{figure}

\begin{figure}
    \centering
    \includegraphics{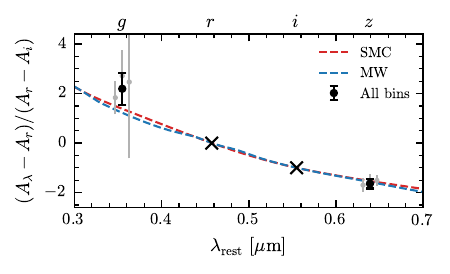}
    \caption{
        Observed extinction relative to the $r$ band, normalized by the $E(r-i)$ of each radial bin to remove the overall extinction amplitude.
        The $i$-band point is therefore fixed along with the $r$-band point.
        For the $g$ and $z$ points, gray shows the individual radial bins and black shows their weighted average.
        The 40--120\,kpc bin is omitted from the gray points because of its low signal-to-noise in $E(r-i)$, but is retained in the weighted average.
        The red and blue curves are the same extinction models as in Fig.~\ref{fig:chromaticity}.
    }
    \label{fig:chromaticity-shape}
\end{figure}

\begin{figure*}[t]
    \centering
    \includegraphics{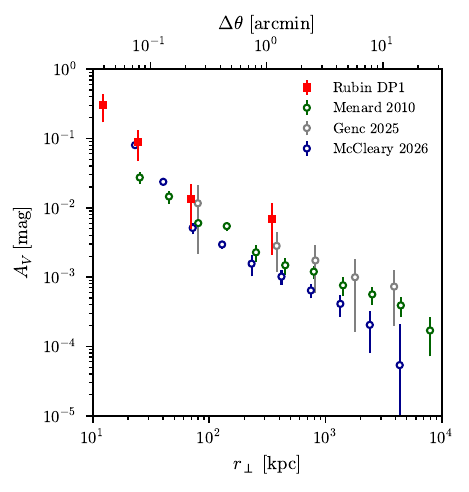}
    \includegraphics{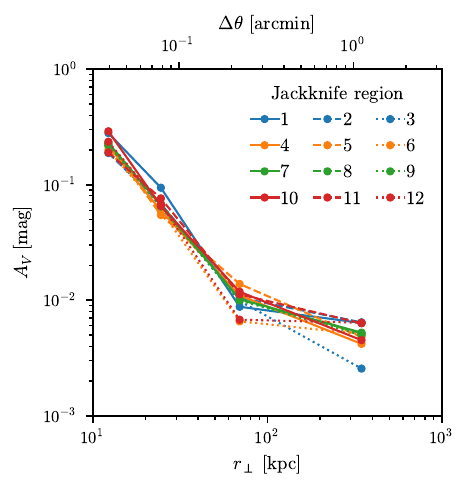}\\
    \includegraphics{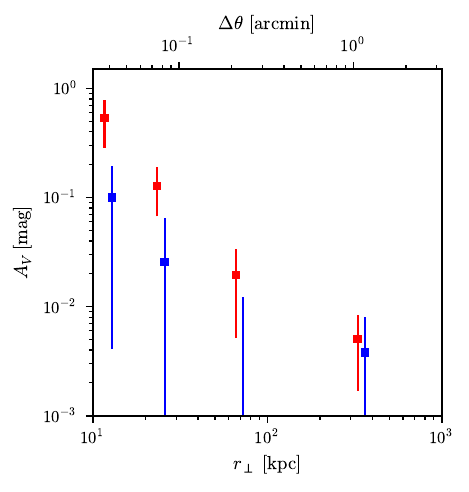}
    \includegraphics{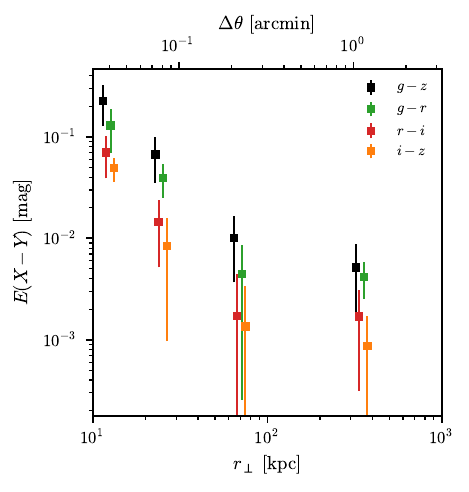}
    \caption{
        Top left: average projected $A_V$ dust extinction inferred from our stacking analysis using Rubin DP1 $E(g-z)$ reddening, compared to previous measurements from SDSS \citep{menard2010}, KiDS \citep{genc2025}, and DES (\citealt{mccleary2026}; WISExSuperCOSMOS sample).
        Top right: $A_V$ measurements from each of the 12 jackknife samples, where the legend indicates which jackknife region was excluded.
        Regions 1--3 are in ECDFS; regions 4--6 are in EDFS; regions 7--9 are in SV~38~7; regions 10--12 are in SV~95~-25.
        Bottom left: $A_V$ for the red and blue foreground samples (cf. Fig.~\ref{fig:foreground}).
        Bottom right: measured color excesses $E(g-z)$, $E(g-r)$, $E(r-i)$, and $E(i-z)$, shown directly in magnitude units.
        Error bars in the top-left and bottom panels show $\pm1\sigma$, and in the bottom panels, points are offset slightly in $r_\perp$ for visibility.
    }
    \label{fig:result}
\end{figure*}

We first test whether the measured color changes have the wavelength dependence expected for dust.
Figure~\ref{fig:chromaticity} recasts the measured $griz$ color excesses in each radial bin as extinction relative to the $r$ band, and compares them with Milky Way and SMC extinction curves.
For each radial bin and extinction law, we fit only the overall normalization.
The resulting $\chi^2$ values indicate that both standard curves provide an acceptable description of the measured chromaticity.
The marginal error bars in Fig.~\ref{fig:chromaticity}, however, are dominated by uncertainty in the common extinction amplitude, which is correlated across bands and therefore does not directly reflect the uncertainty in the wavelength dependence of the signal.
To isolate the wavelength dependence, Fig.~\ref{fig:chromaticity-shape} normalizes each radial bin by its own $E(r-i)$.
In this amplitude-independent comparison, the measurements remain consistent with both extinction-curve shapes.
Given these uncertainties and the limited $griz$ baseline, we do not yet have the leverage to distinguish Milky Way-like from SMC-like extinction.
We adopt the \citet{gordon2023} Milky Way curve with $R_V=3.1$ for the rest of our analyses.

We then estimate the radial profile of the visual extinction.
Because the DP1 footprint provides only 12 jackknife regions, we avoid fitting a full multicolor extinction model that would require inverting a noisy covariance matrix.
Instead, we use $E(g-z)$ as the fiducial color excess and convert it to $A_V$:
\begin{align}
    A_V = \frac{E(g-z)}{(A_g/A_V) - (A_z/A_V)}.
\end{align}
This choice uses the color with the longest wavelength baseline and the highest signal-to-noise.
The band extinction ratios are evaluated through the ComCam bandpasses \citep{comcam_bandpasses}.

$A_V$ profiles are shown in Fig.~\ref{fig:result}.
The fiducial profile in the top-left panel is positive in all four radial bins from $r_\perp \simeq 10$ to $1000$\,kpc and declines smoothly with projected separation, spanning $\sim 0.3$\,mag on the smallest scales to $\sim 0.01$\,mag near $1$\,Mpc.
This panel compares our fiducial curve to the SDSS measurement of \citet{menard2010}, the KiDS measurement of \citet{genc2025}, and the DES measurement of \citet{mccleary2026} for their WISExSuperCOSMOS sample.
Despite the fact that the DP1 area of 4.6\,deg$^2$ is $\sim 1000\times$ smaller than these surveys (see comparison in Table~\ref{tab:sample-comparison}), the Rubin amplitude and slope are consistent with these wide-field measurements over the overlapping radial range.

The top-right panel shows the 12 leave-one-out jackknife profiles that are used to estimate the signal covariance.
Each profile follows similar radial trends, indicating that no single DP1 subregion drives the measurement, except perhaps in our largest radial bin, where the leave-out-region-3 sample exhibits less reddening than the other samples.
This suggests jackknife region 3 (in ECDFS) is responsible for lifting our outermost bin above the collection of other measurements (cf.\ the corresponding point in the top-left panel).
This excess, however, is not significant given the limited data volume in Rubin DP1.

We perform a power-law fit\footnote{
    Because the number of jackknife regions is small, we conservatively neglect the off-diagonal components of the jackknife covariance when fitting the power law, inflating the uncertainty on the power-law exponent by 20\%.
} to the DP1 profile within $120$\,kpc, yielding
\begin{align}
    A_V = (1.3 \pm 0.4) \times 10^{-1} \left( \frac{r_\perp}{20\,\mathrm{kpc}} \right)^{-1.8 \pm 0.4},
\end{align}
indicating a robust detection of a steeply declining projected galaxy--dust correlation with a slope close to the steep inner profile around WISExSuperCOSMOS galaxies found by \citet{mccleary2026}.

The lower-left panel splits the foreground sample at rest-frame $g-r=0.5$.
Power-law fits to the two subsamples over $r_\perp \leq 120$\,kpc give
\begin{align}
    A_V^\mathrm{red} &= (2.0 \pm 0.6) \times 10^{-1} \left( \frac{r_\perp}{20\,\mathrm{kpc}} \right)^{-1.9 \pm 0.5},\\
    A_V^\mathrm{blue} &= (3.2 \pm 4.0) \times 10^{-2} \left( \frac{r_\perp}{20\,\mathrm{kpc}} \right)^{-2.4 \pm 2.7}.
\end{align}
The red-sample profile is higher in the inner CGM, with a best-fit normalization about 6 times larger than that of the blue sample.
The blue-sample normalization, however, is consistent with zero, and its slope is poorly constrained.
Because the split also segregates by stellar and halo mass (Fig.~\ref{fig:foreground}), this is not a clean star-forming/quiescent comparison.

The lower-right panel shows the measured $E(g-z)$, $E(g-r)$, $E(r-i)$, and $E(i-z)$ signals directly.
These color excesses show the same ordering highlighted in Fig.~\ref{fig:chromaticity}: colors involving the bluer $g$ band carry most of the signal, while redder color combinations have smaller excesses.

Our innermost radial bin spans 10--15\,kpc, or 1.7--3.9\,arcseconds, which is a smaller separation than probed in any of the previous surveys to which we compare.
While the smaller stellar masses of our galaxy sample relative to previous measurements suggest that smaller $r_\perp$ may be achievable with our sample, we also note that any diffuse contamination from stars in the foreground will counter reddening from dust, as the low-redshift foreground galaxies are systematically bluer than the high-redshift background galaxies.
For foreground light to substantially bias the $\simeq 0.3$\,mag reddening measured in this bin would additionally require a high level of contamination, which visual inspection and the additional tests in Appendix~\ref{app:inner-bin} suggest is unlikely.

\begin{table*}[t]
    \centering
    \caption{
        Comparison of samples and measurements used to constrain circumgalactic dust reddening.
        DP1 corresponds to this analysis, split into the full foreground sample and the red subsample.
        Other columns are from \citet{menard2010}, \citet{peek2015}, \citet{genc2025}, and the two foreground samples of \citet{mccleary2026}, WISExSuperCOSMOS (W$\times$SC) galaxies and redMaGiC LRGs.
        \emph{Samples:}
        $A$ is survey area (deg$^2$);
        $n_\mathrm{fg}$ and $n_\mathrm{bg}$ are foreground and background number densities (deg$^{-2}$);
        $r_\mathrm{lim}$ is the foreground $r$-band magnitude limit, except where noted;
        $z_\mathrm{fg}$ is the typical foreground redshift: the sample median for DP1, and the sample mean as reported by each of the other works, except \citet{peek2015}, for whom we quote their stated $z\sim0.05$;
        $M_h$ is the typical foreground halo mass ($M_\odot$): the sample median from the \citet{moster2013} relation for DP1, a halo-mass-function-weighted mean for \citet{genc2025}, and an $M_{200m}$ averaged over the halo occupation distribution for the redMaGiC LRGs \citep{zacharegkas2022}.
        \emph{Measurements:}
        the fit range is the interval over which the quoted power-law index is fit, and is narrower than the full data range for DP1 and \citet{mccleary2026};
        the power-law index is the exponent in $A_V \propto r_\perp^{\,\mathrm{index}}$, and for \citet{mccleary2026} describes only their fit range --- their inner W$\times$SC points rise more steeply (cf. Fig.~\ref{fig:result});
        $A_V(150\,\mathrm{kpc})$ is that power law evaluated at $150$\,kpc $\simeq 100\,h^{-1}$\,kpc, the pivot at which \citet{menard2010} and \citet{mccleary2026} report amplitudes;
        $M_\mathrm{dust}$ is the integrated circumgalactic dust mass.
        Dashes mark quantities not reported or not fit.
        \label{tab:sample-comparison}
    }
    \footnotesize
    \setlength{\tabcolsep}{6pt}
    \begin{tabular}{rccccccc}
        \hline\hline
                                          & \multicolumn{2}{c}{DP1 (this work)}                                              &                     &                          &       & \multicolumn{2}{c}{McCleary} \\
        \cline{2-3} \cline{7-8}
                                          &                           all &                             red &         Menard &                Peek &                     Genc &     W$\times$SC &                LRG \\
        \hline
        \multicolumn{8}{l}{\emph{Samples}} \\
        Background tracer                 &                      galaxies &                        galaxies &        quasars &  quiescent galaxies &                 galaxies &            LRGs &               LRGs \\
        $\log_{10}A$                      &                          0.66 &                            0.66 &            3.6 &                 3.9 &                      3.0 &             3.6 &                3.6 \\
        $\log_{10} n_\mathrm{fg}$         &                           3.6 &                             3.3 &            3.8 &                 1.9 &                      2.6 &             2.5 &                2.2 \\
        $\log_{10} n_\mathrm{bg}$         &                           3.6 &                             3.6 &            1.3 &                 1.2 &                      3.6 &             2.6 &                2.3 \\
        $r_\mathrm{lim}$                  &                            24 &                              24 &     21$^{\!*}$ &               17.77 &                       20 &   19.5$^{\!*}$ &                --- \\
        $z_\mathrm{fg}$                   &                          0.36 &                            0.37 &           0.36 &                0.05 &                     0.30 &            0.14 &               0.35 \\
        $\log_{10}M_h$                    &                          11.3 &                            11.7 &            --- &                 --- &  $12.33^{+0.08}_{-0.04}$ &             --- &               13.6 \\
        \hline
        \multicolumn{8}{l}{\emph{Measurements}} \\
        Fit range (kpc)                   &                       10--120 &                         10--120 &      30--$3\times10^4$ &                 --- &                      --- &        74--1500 & 74--$4.5\times10^4$ \\
        Power-law index                   &               $-1.8\pm0.4$    &                    $-1.9\pm0.5$ & $-0.86\pm0.19$ &                 --- &         $-0.8^{\dagger}$ &  $-0.87\pm0.04$ &      $-0.77\pm0.04$ \\
        $A_V(150\,\mathrm{kpc})/10^{-3}$  & $3.2^{+4.5\,\ddagger}_{-1.9}$ &  $3.9^{+7.3\,\ddagger}_{-2.5}$  &    $4.4\pm1.1$ &           $3 \pm 1$ &          $\sim 6^{\,\|}$ &    $3.10\pm0.06$ &         $4.6\pm0.2$ \\
        $M_\mathrm{dust}^{\,\S}/10^{7}M_\odot$ &           $15^{+12}_{-7}$ &                $20^{+21}_{-10}$ &        $\sim5$ &           $6 \pm 2$ &      $7.2^{+0.5}_{-0.6}$ &             --- &                 --- \\
        \hline
        \tabnote{\vspace{2pt}$^{*}$ $i$-band limit for \citet{menard2010}; SuperCOSMOS $R_F$ limit for W$\times$SC.} \\[2pt]
        \tabnote{$^{\dagger}$Fixed rather than fit.} \\[2pt]
        \tabnote{$^{\ddagger}$Values extrapolated from the power-law fit.} \\[2pt]
        \tabnote{$^{\|}$Interpolated from the measured profile of \citet{genc2025} rather than from a power-law fit.} \\[2pt]
        \tabnote{$^{\S}$Integration ranges differ: $20$--$110\,h^{-1}$\,kpc for DP1 and \citet{menard2010}, the nearly identical $30$--$150$\,kpc for \citet{peek2015}, and $20\,h^{-1}$\,kpc--$R_\mathrm{vir} \approx 210\,h^{-1}$\,kpc for \citet{genc2025}; rescaling the latter to a $110\,h^{-1}$\,kpc outer radius gives $\approx 3 \times 10^7\,M_\odot$.
        }
    \end{tabular}
\end{table*}

\section{Discussion}
\label{sec:discussion}

\subsection{What the projected profile traces}

The measured profile is a projected galaxy--dust correlation, not a direct image of dust bound to individual galaxies.
Well inside the virial radius, the signal is naturally associated with a galaxy's own halo; near and beyond the virial radius, it includes two-halo contributions from correlated galaxies, groups, filaments, and diffuse IGM dust.
We fit our power law to $r_\perp \leq 120$\,kpc, which is dominated by the 1-halo term.

The detection therefore reinforces the basic conclusion of earlier work, now for a smaller stellar/halo-mass foreground sample: solid-phase metals have been transported out of the dense ISM and either survive in the CGM or are continuously replenished.

\subsection{The steep inner-dust profile}
\label{sec:steeper}

The slope of our measured reddening profile at $r_\perp < 120$\,kpc, $\sim r_\perp^{-1.8}$, is steeper than the $\sim r_\perp^{-0.8}$ profile measured by \citet{menard2010} on larger scales (cf. Table~\ref{tab:sample-comparison}).
This steep inner slope, however, closely matches the reddening profile around WISExSuperCOSMOS galaxies measured by \citet{mccleary2026}, which they attribute to the influence of active star formation.
Several studies have additionally found evidence for a broken power law with enhanced metal absorption around star-forming galaxies at $r_\perp < 120$\,kpc \citep[e.g.,][]{lan2018}.
Indeed, \citet{chen2025} find a similar $\text{EW(Mg~{\sc ii})} \propto r_\perp^{-1.8}$ profile around emission-line galaxies with similar stellar masses.

The steepness is more striking when interpreted as a three-dimensional dust distribution.
For a spherical power-law density profile, $\rho_\mathrm{dust}(r) \propto r^{-\gamma}$, projecting along the line of sight using the Abel transform yields the projected surface density:
\begin{equation}
    \Sigma_\mathrm{dust}(r_\perp)
    = 2\int_{r_\perp}^\infty \rho_\mathrm{dust}(r)\,\frac{r\,dr}{\sqrt{r^2 - r_\perp^2}}
    \propto r_\perp^{1-\gamma}.
    \label{eq:proj-density}
\end{equation}
Under these assumptions, the observed projected profile corresponds to $\rho_\mathrm{dust}\propto r^{-2.8}$.
This is steeper than both the isothermal $\rho_g\propto r^{-2}$ halo-gas profiles often adopted in cooling models \citep{white1991,maller2004} and the shallower cool-CGM profile inferred by COS-Halos observations \citep{werk2014}.
This suggests the inner-CGM dust distribution is not simply tracing the bulk gas, but instead may be centrally weighted by recent injection from the galaxy, preferentially destroyed or diluted at larger radii, or both.

The amplitude of the innermost bin, $A_V \simeq 0.3$\,mag at $r_\perp=10$--$15$\,kpc, is also notable.
This level of extinction is within a factor of a few of the total high-latitude extinction through the Milky Way disk near the solar circle \citep{lenz2017}.
This is a substantial dust column for a random circumgalactic sight line, and suggests that the small-scale signal is probing either very efficient recent dust injection or long-lived dust close to the galaxy.

\subsection{Implications of consistency across sample depth}
\label{sec:lower-mass}

The agreement with earlier profiles is notable because our DP1 galaxy sample reaches much fainter foreground magnitudes and lower inferred stellar masses than previous wide-field measurements.
Our selection extends to $r<24$ and stellar masses of order $10^8\,M_\odot$, while previous foreground samples were dominated by brighter galaxies with minimum stellar masses closer to $10^9$--$10^{10}\,M_\odot$ (Table~\ref{tab:sample-comparison}; Fig.~\ref{fig:foreground}).
However, Fig.~\ref{fig:foreground} and the color split in Fig.~\ref{fig:result} show that the lowest-mass galaxies are concentrated in the blue subsample, whose reddening signal is detected at low significance.
The full-sample agreement therefore should not be read as a direct detection of circumgalactic dust around the lowest-mass DP1 galaxies.

It instead shows that extending the foreground selection to fainter and lower-mass galaxies does not erase the population-averaged reddening signal.
This remains consistent with scenarios in which lower-mass galaxies contribute efficiently to circumgalactic dust per unit stellar mass because their shallower potentials and bursty feedback eject a larger fraction of their metals and dust into the CGM \citep{chisholm2017,chisholm2018,zheng2024,martin-alvarez2026}, but the larger samples offered by LSST will be needed to test this trend directly.

\subsection{CGM dust around lower-mass red galaxies}
\label{sec:red-galaxies}

The lower-left panel of Fig.~\ref{fig:result} suggests that at $r_\perp < 120$\,kpc the red-sample profile is higher than the blue-sample profile, although the blue fit is too poorly constrained to establish a significant red-blue difference.
The red-sample signal is nevertheless in contrast to the results of \citet{mccleary2026}, who find little reddening in the inner CGM of redMaGiC LRGs despite substantial reddening on scales above 100\,kpc.

This contrast may be explained by the fact that LRGs occupy more massive halos that have been evacuated by feedback.
The redMaGiC sample has characteristic halo mass $\approx 5 \times 10^{13}\,M_\odot$ \citep{zacharegkas2022}, and recent kSZ measurements of LRGs favor ionized-gas profiles that are more extended than the dark matter, consistent with strong feedback redistributing baryons out of the inner halo \citep{riedguachalla2025}.
In such systems, a weak inner-dust signal may reflect both a reduced cool gas column in the inner CGM and efficient destruction of any remaining grains in hot halo gas.

Our red galaxy sample is much less massive: its median halo mass is $\approx 5 \times 10^{11}\,M_\odot$, 2 orders of magnitude smaller than the LRGs.
Because $T_\text{vir} \propto M_\text{vir}^{2/3}$ \citep{barkana2001}, our red sample has a virial temperature $\sim 20\times$ lower, around $5 \times 10^5$\,K, compared with $\sim 10^7$\,K for redMaGiC LRGs.
This lies below the $\sim10^6$\,K threshold where sputtering becomes efficient in the CGM \citep{richie2024}, so dust may survive for long durations in many of our red-sample halos.

We note that a simple rest-frame color cut does not isolate a uniformly passive population, and may include dusty star-forming galaxies and inclined disks.
The current DP1 samples are too small to disentangle color, stellar mass, star-formation activity, morphology, and environment, but our results suggest that the LRG inner-CGM dust deficit should not be generalized to lower-mass red galaxies.

\subsection{Dust mass and metal budget}

We estimate a halo dust mass for the red subsample by interpreting $A_V$ as a projected optical-depth profile:
\begin{align}
    A_V(r_\perp) = 1.086 \, \tau_V(r_\perp) \equiv 1.086 \, \kappa_V \Sigma_\text{dust}(r_\perp),
\end{align}
where $\kappa_V$ is the dust mass opacity and $\Sigma_\text{dust}$ is the population-weighted, azimuthally averaged projected dust density.
Asymmetries in the dust distribution of individual halos average out in the stack, increasing the variance of the estimate without biasing it.
The mean dust mass is then given by integrating the profile over an annulus:
\begin{align}
    M_\text{dust}^\text{CGM}
    = 2 \pi \int_{r_\text{inner}}^{r_\text{outer}} \Sigma_\text{dust}(r_\perp) \, r_\perp dr_\perp.
\end{align}
Following \citet{menard2010}, we assume $\kappa_V \approx 2 \times 10^4$\,cm$^2$\,g$^{-1}$ and adopt $(r_\text{inner}, r_\text{outer}) = (20, 110)\,h^{-1}$\,kpc.
Accounting for uncertainties in the red-sample power-law fit, we find
\begin{align}
    M_\text{dust}^\text{CGM} \approx 2.0^{\,+2.1}_{\,-1.0} \times 10^8 \, M_\odot.
\end{align}
Note that estimating dust masses with LMC- or SMC-like assumptions yields values consistent within the error bars.
Repeating this calculation for the full foreground sample gives $1.5^{\,+1.2}_{\,-0.7} \times 10^8 \, M_\odot$ (Table~\ref{tab:sample-comparison}).

This estimate is about 4 times larger than the \citet{menard2010} estimate of $M_\text{dust}^\text{CGM} \approx 5 \times 10^7\,M_\odot$ for their $\sim0.5L^\star$ population.
This larger mass is not unexpected given the shape of the profile:
\citet{menard2010} measured a shallower $\sim r_\perp^{-0.8}$ profile over larger scales, whereas the DP1 red sample and the WISExSuperCOSMOS sample of \citet{mccleary2026} show a much steeper inner-CGM component (see Section~\ref{sec:steeper}).

Normalizing by the red-sample median stellar and halo masses yields the dust-to-mass ratios
\begin{align}
    \frac{M_\text{dust}^\text{CGM}}{M_\star} &\approx 1.8^{\,+1.9}_{\,-0.9} \times 10^{-2} \\
    \frac{M_\text{dust}^\text{CGM}}{M_h} &\approx 4.1^{\,+4.3}_{\,-2.0} \times 10^{-4}.
\end{align}
These values are roughly an order of magnitude greater than the values reported by \citet{menard2010,genc2025}.
Because our red sample extends to stellar masses roughly an order of magnitude below these samples (cf. Fig.~\ref{fig:foreground}), this result may be consistent with scenarios in which lower-mass galaxies contribute efficiently to circumgalactic dust per unit stellar mass (see Section~\ref{sec:lower-mass}).

These ratios are comparable to the total dust budget expected from simple metal-yield arguments.
For a standard IMF, stellar populations return a metal mass of order $y_Z M_\star$, with $y_Z \sim 0.01$--$0.03$ \citep{vincenzo2016}.
If an L$^\star$-like fraction $f_\mathrm{dust/Z}\sim0.3$--$0.6$ of those metals resides in grains \citep{peeples2014}, the implied dust budget is $\sim 0.003$--$0.02\,M_\star$.
Our estimated mass of $0.018\,M_\star$ nearly saturates this limit.

These ratios should not be overinterpreted beyond an order-of-magnitude comparison, owing to uncertainties in our stellar-mass estimates, the dust opacity, stellar-mass weighting, and the mass/metallicity dependence of the dust-to-metal ratio \citep{remyruyer2015,devis2019}.
Our result, however, suggests the CGM may act as a reservoir for a large fraction of the available solid-phase metal budget.

\subsection{Prospects for LSST}
\label{sec:lsst}

DP1 is a pathfinder for LSST, which will provide $4000\times$ more sky area and dramatically reduce the jackknife noise floor that limits this analysis, making it possible to constrain the galaxy--dust correlation on much larger distance scales.
This will enable separating the one- and two-halo terms, testing for departures from a single power law, and measuring how the profile evolves with redshift.

The larger sample will also make galaxy--dust reddening a genuinely differential probe of the baryon cycle.
LSST will support simultaneous binning by stellar mass, star-formation rate, morphology, inclination, and environment.
LSST's full six-band photometry will constrain the wavelength dependence of the reddening signal, and thereby distinguish changes in dust column from changes in grain size population.
This will help determine whether CGM dust in low-mass halos is more similar to Milky Way, LMC, or SMC extinction.

Precision large-scale galaxy--dust correlations will also be useful for cosmological studies, informing Type~Ia supernova extinction systematics, as well as magnification and clustering analyses.

\section{Conclusion}
\label{sec:conclusion}

We have presented the first circumgalactic dust-reddening measurement from the Rubin Observatory.
Using only $4.6$\,deg$^2$ of DP1 commissioning data, we detect a positive, smoothly declining, chromatic projected-reddening profile across four radial bins from $r_\perp\simeq 10$\,kpc to $1$\,Mpc.
The signal is stable across jackknife regions and analysis variants, and the wavelength dependence is consistent with both Milky Way and SMC-like dust extinction.

Interpreted as a spherical density profile, the projected slope implies $\rho_\mathrm{dust}\propto r^{-2.8}$, steeper than common halo-gas profiles and suggesting that inner-CGM dust does not simply trace the bulk gas.
The innermost bin reaches $A_V\simeq0.3$\,mag at $r_\perp=10$--$15$\,kpc, comparable to the high-latitude Milky Way disk column near the solar circle.
Visual inspection and blendedness tests indicate that foreground light contamination is unlikely to significantly bias this signal.

Despite covering far less sky area than earlier SDSS, KiDS, and DES studies \citep{menard2010,peek2015,genc2025,mccleary2026}, Rubin's depth and high source densities yield a profile with comparable amplitude and radial dependence.
This agreement is especially informative because the DP1 foreground selection reaches $r<24$, 3--6\,mag fainter than previous studies, and extends to stellar masses as low as $\sim10^8\,M_\odot$, 1--2\,dex lower than earlier samples.

A split by rest-frame $g-r$ color shows that our red sample (rest-frame $g-r > 0.5$) has a best-fit inner-CGM normalization $\approx 6\times$ higher than the blue sample, although the blue sample is currently too noisy to draw definite conclusions.
Because the lowest-mass foregrounds are concentrated in the noisy blue subsample, the present data should be read as a robust population-averaged detection rather than as a direct measurement of the dust content of $M_\star \lesssim 10^9\,M_\odot$ galaxies.

Our red galaxy sample, with median halo mass $5 \times 10^{11}\,M_\odot$, exhibits substantially more dust reddening within 50\,kpc than has previously been measured around more massive LRG populations \citep{mccleary2026}.
This comparison suggests the LRG inner-dust deficit should not be generalized to lower-mass red galaxies, whose halos are cooler, less feedback-evacuated, and may include dusty star-forming systems that replenish circumgalactic dust.

For the red sample, integrating the fitted profile over $20$--$110\,h^{-1}$\,kpc gives $M_\mathrm{dust}^\mathrm{CGM}\approx2\times10^8\,M_\odot$.
Normalized by the median stellar and halo masses, this corresponds to $M_\mathrm{dust}^\mathrm{CGM}/M_\star\approx1.8\times10^{-2}$ and $M_\mathrm{dust}^\mathrm{CGM}/M_h\approx4.1\times10^{-4}$, about an order of magnitude above estimates for the more massive galaxy samples of \citet{menard2010,genc2025}.
These high ratios are consistent with lower-mass galaxies enriching their CGM efficiently per unit stellar mass.
The dust-to-stellar-mass ratio also nearly saturates the simple dust budget allowed by stellar metal yields, suggesting the CGM may contain a large fraction of the available solid-phase metals in these galaxies.

Our DP1 analysis provides a compelling pathfinder for LSST.
The full survey will transform this first detection into precise measurements as a function of galaxy mass, star-formation activity, environment, and redshift, and will enable joint analyses with weak lensing, absorption-line probes, SZ gas tracers, and FRB dispersion measures.
Together, these measurements will connect solid-phase metals to the broader baryon cycle and help calibrate extragalactic dust systematics for next-generation cosmology.

\vspace{-2mm}
\begin{acknowledgments}
    We are grateful to Sergio Martin-Alvarez for detailed feedback on the manuscript, and to Viraj Manwadkar, Susan Clark, Risa Wechsler, Rebecca Chen, and Erik Peterson for useful discussions.

    This project began during a sprint week hosted by the DiRAC Institute in the Department of Astronomy at the University of Washington.
    The DiRAC Institute is supported through generous gifts from the Charles and Lisa Simonyi Fund for Arts and Sciences, Janet and Lloyd Frink, and the Washington Research Foundation.

    J.F.C. acknowledges support from the U.S. Department of Energy, Office of Science, Office of High Energy Physics Cosmic Frontier research program under award number DE-SC0022083.
    J.K.W. gratefully acknowledges support from NSF-CAREER 2044303.

    This material is based upon work supported in part by the National Science Foundation through Cooperative Agreements AST-1258333 and AST-2241526 and Cooperative Support Agreements AST-1202910 and 2211468 managed by the Association of Universities for Research in Astronomy (AURA), and the Department of Energy under Contract No. DE-AC02-76SF00515 with the SLAC National Accelerator Laboratory managed by Stanford University. Additional Rubin Observatory funding comes from private donations, grants to universities, and in-kind support from LSST-DA Institutional Members.

    This publication is based in part on proprietary Rubin Observatory Legacy Survey of Space and Time (LSST) data, and was prepared in accordance with the Rubin Observatory data rights and access policies. All authors of this publication meet the requirements for coauthorship of proprietary LSST data.

    This research uses services or data provided by the Rubin Science Platform at NSF-DOE Vera C. Rubin Observatory, which is jointly funded by the U.S. National Science Foundation and the U.S. Department of Energy, Office of Science.

    \texttt{Claude Sonnet 4.6} and \texttt{GPT-5.5} were used for some of the coding in this project, especially for refactoring and improving computational efficiency, as well as for reviewing the manuscript text for grammar and clarity.
    The authors take full responsibility for the scientific results and the written text.
\end{acknowledgments}

\facilities{
    Rubin:Simonyi (LSSTComCam),
    Rubin:USDAC
}

\software{
    \texttt{astropy} \citep{astropy:2013,astropy:2018,astropy:2022},
    \texttt{dust\_extinction} \citep{gordon2024},
    \texttt{healpy} \citep{zonca2019,gorski2005},
    \texttt{jupyter} \citep{jupyter},
    \texttt{kcorrect} \citep{blanton2007},
    \texttt{matplotlib} \citep{matplotlib},
    \texttt{numpy} \citep{numpy},
    \texttt{pandas} \citep{pandas,pandas-software},
    \texttt{scipy} \citep{scipy},
    \texttt{seaborn} \citep{seaborn},
    \texttt{treecorr} \citep{jarvis2004},
    \texttt{Claude Sonnet 4.6},
    \texttt{GPT-5.5}
}

\appendix

\vspace{-6mm}
\section{Random-catalog construction}
\label{app:randoms}

The fiducial random correction uses random catalogs drawn from the same selected HEALPix footprint as the galaxy sample, with random positions sampled uniformly within each jackknife region.
Because the galaxy selection varies with imaging depth, we reweight the random catalogs to match the local depth distribution of the corresponding real sample.

For the foreground and background samples separately, in each HEALPix pixel we compute the median $5\sigma$ depth in each of the $griz$ bands, which we denote $g_5,\,r_5,\,i_5,\,z_5$.
Within each jackknife region, we divide both the real galaxies and randoms into three quantile bins in each depth dimension, forming $3^4=81$ joint cells in $(g_5,r_5,i_5,z_5)$ depth space for each sample.
Random points are then weighted by cell using the ratio
\begin{equation*}
    w_\text{cell} =
    \frac{N^\mathrm{gal}_\text{cell} / N^\mathrm{gal}}
         {N^\mathrm{rand}_\text{cell} / N^\mathrm{rand}}.
\end{equation*}
The weights are normalized to unit mean within each jackknife region so the random catalogs preserve their overall normalization while matching the depth distribution of the corresponding real sample.

For the random foreground catalog, we preserve the real foreground redshift distribution by assigning each random position a foreground-galaxy photo-$z$ drawn from the corresponding jackknife region.
This keeps the conversion from angular separation to $r_\perp$ matched to the data.
We then measure the same forward-minus-flipped estimator around random positions,
\begin{align*}
    \Delta \bar c^\mathrm{rand}_i
    \equiv \bar{c}^\mathrm{rand}_i
    - \bar{c}^{\mathrm{rand,flip}}_i ,
\end{align*}
where $\bar{c}^\mathrm{rand}_i$ stacks background-galaxy colors around random foreground positions and $\bar{c}^{\mathrm{rand,flip}}_i$ stacks foreground-galaxy colors around random background positions.

\section{Sensitivity to analysis choices}
\label{app:sensitivity}

\begin{figure}
    \centering
    \includegraphics{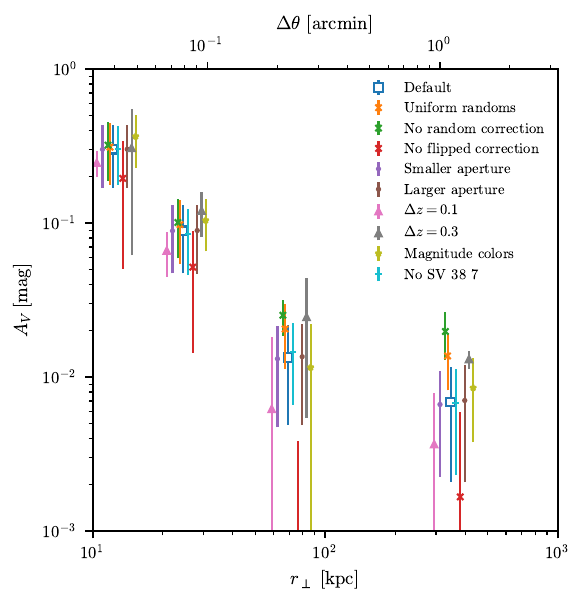}
    \caption{
        Stacked reddening profiles with jackknife uncertainties for several analysis variants.
        With the exception of the estimator that omits the flipped-stack correction, the variants remain consistent with the fiducial measurement, indicating that the DP1 signal is not driven by any single analysis choice.
        All analyses use the same radial binning; the offsets in $r_\perp$ are for visibility.
    }
    \label{fig:sensitivity}
\end{figure}

The projected-reddening estimator described in Section~\ref{sec:methods} combines several corrections: depth-matched random catalogs, a flipped-stack subtraction, a finite reference annulus, and a foreground--background photo-$z$ separation.
Our analysis also combines DP1 fields of varying data quality.
Most notably, SV~38~7 lacks $u$- and $y$-band imaging, degrading its photo-$z$ estimates and potentially shifting its mass and color distributions.
We test whether the measured signal depends sensitively on these choices by recomputing the full stack under the variants shown in Fig.~\ref{fig:sensitivity}.
These variants are:
\begin{itemize}
    \item ``Uniform randoms,'' in which all random positions are assigned equal weight rather than weights matched to the local depth distribution;
    \item ``No random correction,'' in which we do not subtract the estimator measured around random positions;
    \item ``No flipped correction,'' in which we do not subtract the flipped stack of foreground-galaxy colors around background positions;
    \item ``Smaller aperture,'' in which the reference annulus is shifted inward to $1.5 < r_\perp/\mathrm{Mpc} < 3.5$;
    \item ``Larger aperture,'' in which the reference annulus is shifted outward to $2.5 < r_\perp/\mathrm{Mpc} < 4.5$;
    \item ``$\Delta z = 0.1$,'' in which the lower limit of the background sample is moved to $z_\text{phot} = 0.6$;
    \item ``$\Delta z = 0.3$,'' in which the lower limit of the background sample is moved to $z_\text{phot} = 0.8$;
    \item ``Magnitude colors,'' in which the stack uses magnitude differences rather than flux ratios;
    \item ``No SV~38~7,'' in which this DP1 field is excluded from the stacking analysis.
\end{itemize}

Nearly all variants agree with the fiducial profile within the jackknife uncertainties.
This agreement indicates the detection does not rely on the precise random-catalog weighting, the exact placement of the reference annulus, the adopted width of the photo-$z$ gap between the foreground and background samples, or the definition of ``color'' in the stack.
The missing $u$- and $y$-band imaging in SV~38~7 likewise does not measurably bias the profile within the DP1 uncertainties.
The upper-right panel of Fig.~\ref{fig:result} points to the same conclusion: the jackknife samples that exclude subregions of SV~38~7 agree well with the rest.
The notable exception is the ``No flipped correction'' case, which is substantially noisier at all radii and shifts low on large scales.
Because this behavior coincides with increased scatter among the jackknife regions, it suggests that the flipped-stack subtraction is removing spatially coherent photometric or selection residuals that are not fully captured by the random correction alone.
Once that correction is included, the stability of the profile across jackknife regions and across the remaining analysis variants supports the robustness of the DP1 reddening measurement.

\begin{figure}
    \centering
    \includegraphics[width=\linewidth]{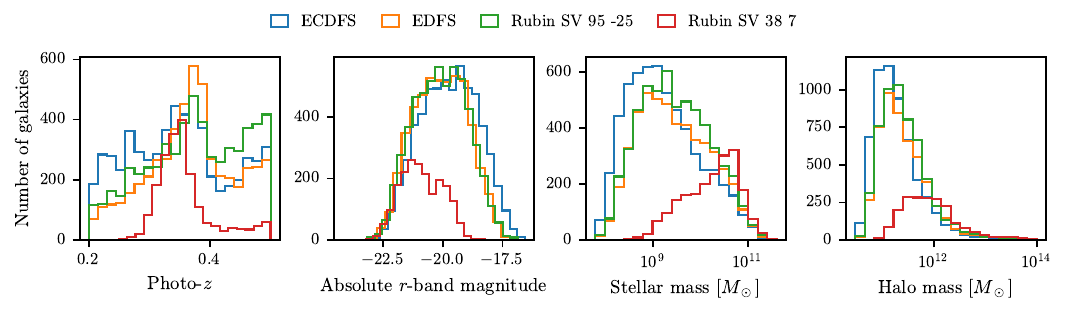}
    \caption{
        Properties of the foreground-galaxy sample, split by DP1 field.
        Left: distribution of photo-$z$ point estimates;
        Middle left: absolute $r$-band magnitudes;
        Middle right: stellar-mass estimates;
        Right: halo mass estimates.
    }
    \label{fig:foreground-by-field}
\end{figure}

We investigate SV~38~7 further by comparing the foreground samples selected in each DP1 field (Fig.~\ref{fig:foreground-by-field}).
Because it is shallower, this field contributes primarily to the bright, high-mass end of the population.
Its lack of $u$-band imaging also leaves it with almost no galaxies below $z \approx 0.3$ where photo-$z$ estimates rely on $u-g$ to bracket the Balmer break (rest-frame 3646\,\AA).
Every field shows a local maximum at $z \approx 0.35$--$0.4$, where the break sits well into the $g$ band and is bracketed by $g-r$, the color formed from the two deepest bands.
The distribution then declines as the break moves across the $g/r$ filter boundary and $g-r$ saturates.
It rises again by $z \approx 0.45$, once the break has moved into $r$ and $r-i$ regains sensitivity.
This second rise is strongly suppressed in SV~38~7, likely because of its shallow $r$- and $i$-band depths.
Despite these differences, the sensitivity test in Fig.~\ref{fig:sensitivity} shows that including SV~38~7 does not measurably change our results within the DP1 uncertainties.

\section{Validating the robustness of the innermost radial bin}
\label{app:inner-bin}

The innermost radial bin in our stack spans 10--15\,kpc in $r_\perp$, or 1.7--3.9\,arcsec in angular separation.
This bin probes a smaller separation than the previous surveys to which we compare, namely \citet{menard2010,peek2015,genc2025,mccleary2026}, and there is potential for significant contamination of the background fluxes by the foreground object.
We therefore perform several tests to evaluate the robustness of the measurement in this innermost bin.

\begin{figure}
    \centering
    \includegraphics{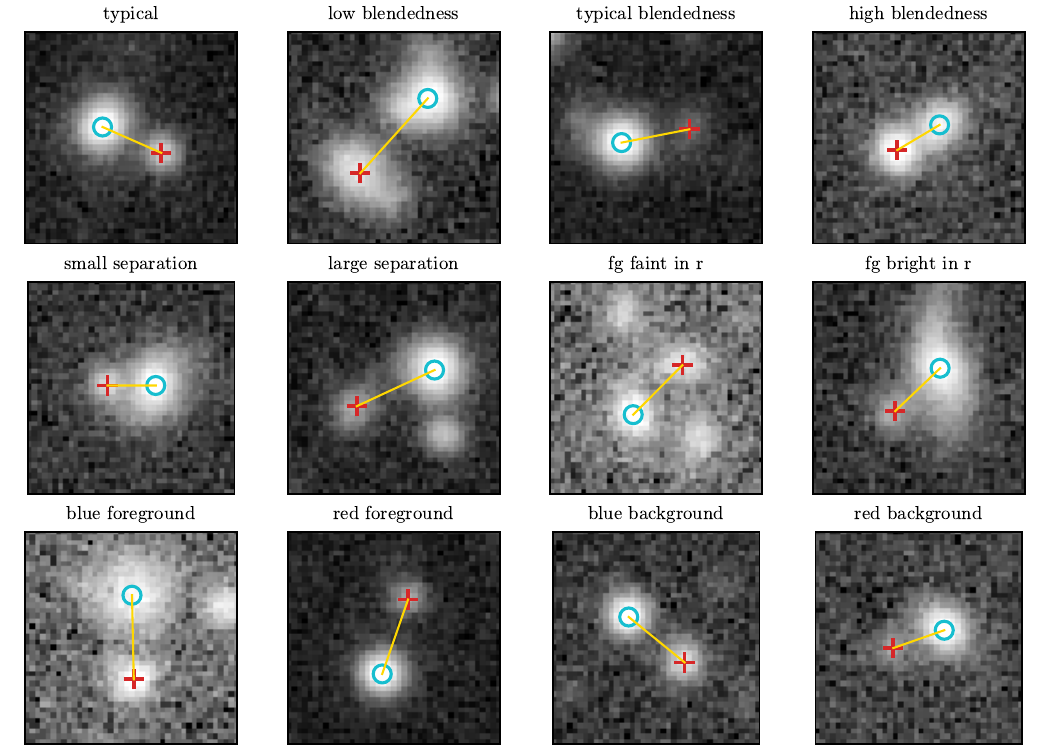}
    \caption{
        A set of representative foreground (blue) and background (red) pairs from the innermost radial bin in Fig.~\ref{fig:result}, spanning different blendedness values, separation scales, foreground fluxes, and foreground/background $g-z$ colors.
        The ``typical'' pair is selected to be very close to the median in all of these quantities.
    }
    \label{fig:stamp-grid}
\end{figure}

First, we define a \emph{blendedness} metric.
For each band, the DP1 catalog provides \texttt{deblend\_fluxOverlap}, which quantifies the total flux from \emph{all} neighboring objects that overlaps the source footprint in the deconvolved Scarlet \citep{melchior2018} deblender model.
\texttt{deblend\_fluxOverlapFraction} divides this neighboring flux by the source flux within the deconvolved footprint.
Our blendedness metric is the maximum of \texttt{deblend\_fluxOverlapFraction} across $griz$.
Note this metric includes blending from any neighboring object, not just the foreground galaxy in the pair.
It is therefore not a direct estimate of how much foreground flux leaked into the background galaxy's measured photometry.
Instead, it is a conservative model-based diagnostic of how much neighboring-source flux is present in the object's deblended footprint.

We then visually inspected all 111 pairs in this bin to determine whether each pair contains visually distinct objects or is heavily blended.
A representative set is shown in Fig.~\ref{fig:stamp-grid}.
These pairs were chosen to span the bin's blendedness values, separations, foreground fluxes, and foreground/background $g-z$ colors.
The upper-left pair was chosen to be very close to the median in these properties.
While some pairs clearly experience some amount of blending (e.g., the pairs labeled ``high deblend overlap'' and ``small separation''), each pair contains two clearly distinct objects.
Moreover, because the reddening in the innermost bin is at the 20\% level, biasing our measurement would demand substantial contamination, far more than these images suggest.
We note that because of their lower redshifts, the foreground galaxies are systematically \emph{bluer} than the background galaxies.
Any blending, therefore, is likely to dilute the stacking signal, rather than strengthen it.

We then perform a sensitivity test, evaluating the amplitude of the stacked colors while splitting the sample by blendedness.
We perform this test on both of the first two radial bins, where the larger bin, which should not experience any significant foreground--background blending, serves as a reference benchmark for comparison.

The split difference in the first (second) bin is $\Delta(g-z) = -0.04\,(-0.001)$, indicating that, as expected, the background galaxies in more highly blended pairs are observed to be systematically bluer.
This represents a $-0.4\sigma\,(-0.05\sigma)$ shift in this bin when compared to the error bar of the full stack.
Determining the uncertainty in $\Delta(g-z)$ via bootstrapping the objects in each split, the shift is only $-0.2\sigma\,(-0.04\sigma)$.
Finally, we perform a two-sided permutation test by randomly shuffling the blendedness labels within each bin many times.
This provides a $p$-value that answers the question ``how often would random labels produce a delta at least this large?''
We find $p=0.82\,(0.97)$ for the first (second) bin.
These tests suggest any blending contamination of the innermost bin is small compared to the amplitude of the signal and its error bar.

The robustness of our sample to blending may result in part from requiring strong agreement between the template-based and machine-learning photo-$z$ estimates (see Section~\ref{sec:photoz}).
Blending degrades photo-$z$ estimation by producing composite fluxes and colors that are not well described by a single-galaxy SED \citep{jones2019}.
Because template-based estimators rely on physical models of galaxy spectra while machine-learning estimators learn empirical mappings from a spectroscopic training set, the two approaches are not expected to fail in the same way in the presence of blending.
The photo-$z$ agreement criterion may therefore reject a substantial fraction of blended contaminants \citep{liang2026}.

Finally, the innermost bin also experiences the greatest lensing magnification from the foreground halos.
This can bias the estimator if the color distribution of the background sample has a strong derivative with $r$-band magnitude.
To estimate the size of this bias, we approximate each foreground halo as an NFW halo \citep{navarro1997} with concentration $c=10.0$ \citep{ludlow2014}, using the halo masses assigned to the DP1 foreground sample and the projected separations of the 10--15\,kpc pairs.
The resulting characteristic magnification is of order 2\%.
The empirical background-sample $d(g-z)/dr \approx -0.4$, yielding $\Delta(g-z) \approx -0.009$.
This is approximately 4\% of the signal amplitude and 9\% of the jackknife uncertainty.
Under these assumptions, magnification-induced color selection is unlikely to significantly bias the observed inner-bin reddening signal.

\bibliographystyle{aasjournalv7.1}
\bibliography{references}

\end{document}